\input{aipcheck}

\documentclass[
    ,final            
  ]
  {aipproc}

\layoutstyle{6x9}

\begin{document}

\title{Broad Absorption Line Quasar catalogues with Supervised Neural Networks}


\author{Simone Scaringi}{
  address={Department of Physics and Astronomy, University of Southampton, Highfield, SO17 1BJ, U.K.}
}

\author{Christopher E. Cottis}{
  address={Department of Physics and Astronomy, University of Leicester, University road, LE1 7RH, U.K.}
}

\author{Christian Knigge}{
  address={Department of Physics and Astronomy, University of Southampton, Highfield, SO17 1BJ, U.K.}
}

\author{Michael R. Goad}{
  address={Department of Physics and Astronomy, University of Leicester, University road, LE1 7RH, U.K.}
}

\begin{abstract}
We have applied a Learning Vector Quantization (LVQ) algorithm to SDSS DR5 quasar spectra in order to create a large catalogue of broad absorption line quasars (BALQSOs). We first discuss the problems with BALQSO catalogues constructed using the conventional balnicity and/or absorption indices (BI and AI), and then describe the supervised LVQ network we have trained to recognise BALQSOs. The resulting BALQSO catalogue should be substantially more robust and complete than BI- or AI-based ones.
\end{abstract}

\maketitle


\section{Introduction}
Broad absorption line quasars (BALQSOs) are a sub-class of active
galactic nuclei (AGN) exhibiting strong, broad and blue-shifted
spectroscopic absorption features
(\cite{foltz90,weymann91,reichard03b,hewett03}) associated with strong winds of outflowing material reaching 0.1$c$-0.2$c$ \cite{korista92} . BALQSOs are predominantly radio-quiet (\cite{stocke92}), and there are subtle differences between their continuum and emission line properties and those of ``normal'' (non-BAL) QSOs (\cite{reichard03b}). However, despite these differences, BALQSOs and non-BALQSOs appear to be drawn from the same parent population (\cite{reichard03b}).

The most straightforward explanation for the differences between QSOs
and BALQSOs is a simple orientation effect. Thus {\em all} QSOs may
undergo significant mass loss through winds, but BALs are only observed if 
the central continuum and/or emission line source is viewed directly
through the outflowing material. Viewed in this context, BALQSOs may
be the only available tracers of a key physical process common to all
AGN. Moreover, the fraction of QSOs displaying BAL features
($f_{BALQSO}$) may provide a direct estimate of the opening angle of
these outflows. 

Historically, most BALQSO samples were selected on the basis of the
so-called balnicity index (BI; \cite{weymann91}) or similar
metrics. These samples consistently yielded BALQSO fraction estimates
in the range $f_{BALQSO} \approx 0.10 - 0.15$ (\cite{weymann91, Tolea,
hewett03, reichard03a}). In a previous paper (\cite{knigge08}; Paper I), we showed that the BI and a recently defined metric, the absorption index (AI; \cite{trump06}), are to strict or relaxed respectively when selecting BALQSOs.

Here we will use the hybrid-LVQ approach from Paper I using a combination of the classic BI, a simple neural network and visual inspection to produce BALQSO samples that are more robust than AI-based ones, but more complete than purely BI-based ones using the QSO sample associated with Data Release 5 (DR5) of the Sloan Digital Sky Survey (SDSS; \cite{BIG_dr5,dr5}). Our catalogue contains 3505 BALQSOs selected from 28,421 objects in the SDSS DR5 QSO sample in the redshift range $1.7<z<4.2$.

\section{The Input QSO Sample}

The SDSS DR5 QSO catalogue contains over 77,000 objects in total
\cite{dr5}. However, for the purpose of constructing a uniform
BALQSO catalogue, we only consider objects whose spectra
fully cover the C~{\sc iv}~1550~\AA\ resonance line, which displays
a particularly deep and well-defined absorption through in the spectra
of most BALQSOs. Given the wavelength range covered by the SDSS
spectra, this implies an effective redshift window of $1.7<z<4.2$ for
our QSO parent sample, which contains spectra of 28,421 objects.

Our BALQSO classification method works on continuum-normalised spectra 
covering the wavelength range 1401 - 1700~\AA\ with 1~\AA\
 dispersion. It also uses the associated BIs for training the
neural network and to flag borderline cases requiring visual
inspection. We therefore normalise all QSO spectra using the method
described in \cite{knigge08, north06}, in which each spectrum is fit
with a modified DR5 QSO composite allowing for object-to-object
differences in reddening and overall spectral slope
\cite{vanden01}. We then interpolate each spectrum onto the new
wavelength grid and estimate the BI in the same way as
\cite{weymann91}.

\section{Hybrid-LVQ Selection of BALQSOs}

\begin{figure}
\centering \includegraphics[width=0.5\textwidth, height=0.35\textheight]{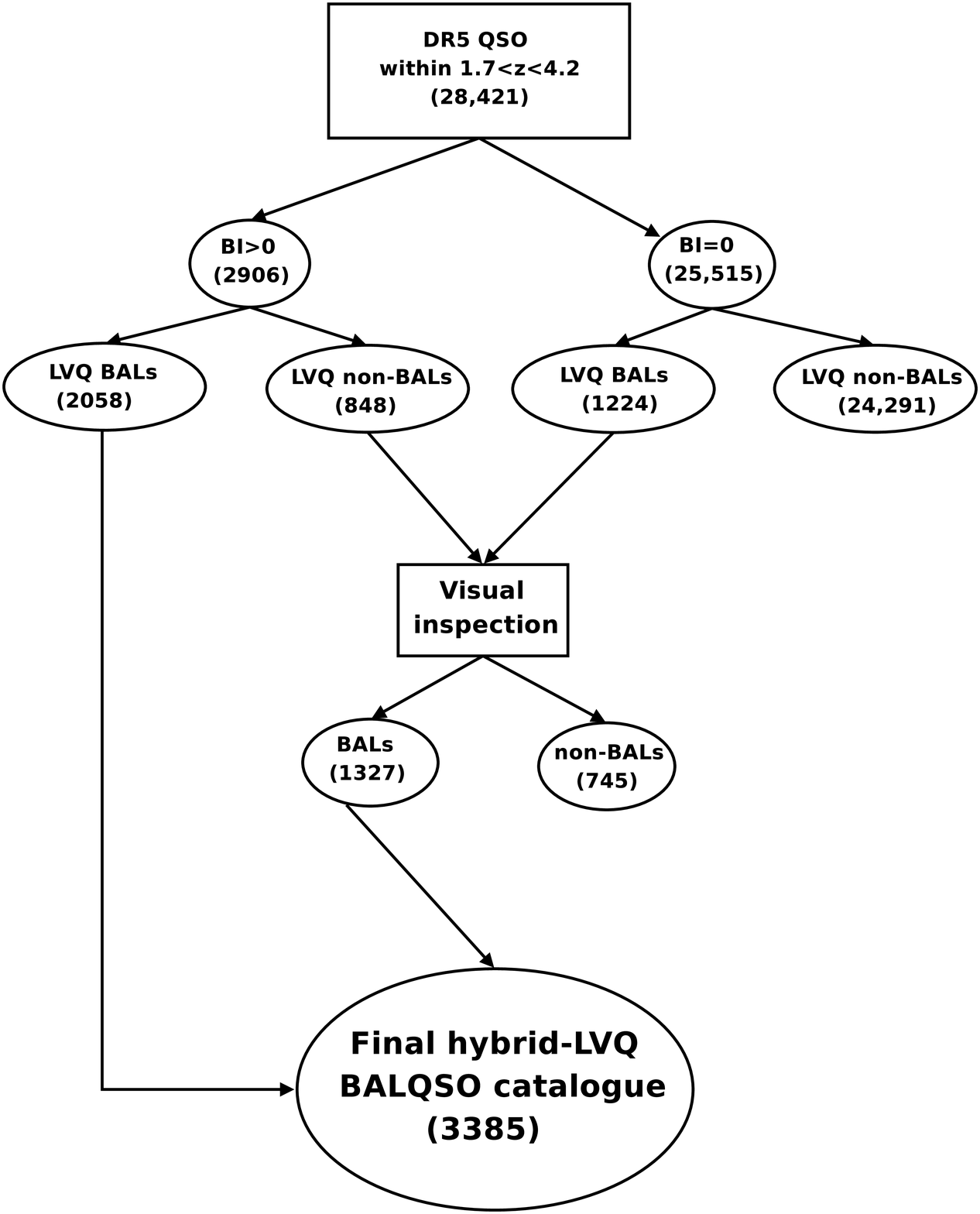}
\caption{Flow diagram illustrating the steps involved in our
hybrid-LVQ classification method.}
\label{fig:dr5_flow}
\end{figure}

The method we use to classify BALQSOs has already been described in
detail in \cite{knigge08}, so we only provide an overview of the key
points here. Briefly, our method is a hybrid of BI-based, neural
network and visual classifications. It is designed to produce a more
complete BALQSO sample than a pure BI selection without significantly
increasing the number of false positives. Starting with a BI-based
classification, we use a simple neural network-based machine learning
algorithm called 
``learning vector quantization'' (LVQ, \cite{kohonen01}) to  
identify objects that might have been misclassified by the BI. All
such objects are then inspected and classified visually. The way in
which we train our LVQ network to recognize BALQSOs has been described 
in detail in Paper~I.  Note that redshift uncertainties are explicitly
taken into account by our network. Below, we will sometimes refer to
the full hybrid method as ``LVQ-based'', but it is always worth
keeping in mind that LVQ is only one part of a process also involving
the BI and visual inspection.

\section{Properties of the Final BALQSO Catalogue}

\begin{figure*}
\centering \includegraphics[width=\textwidth, height=0.25\textheight]{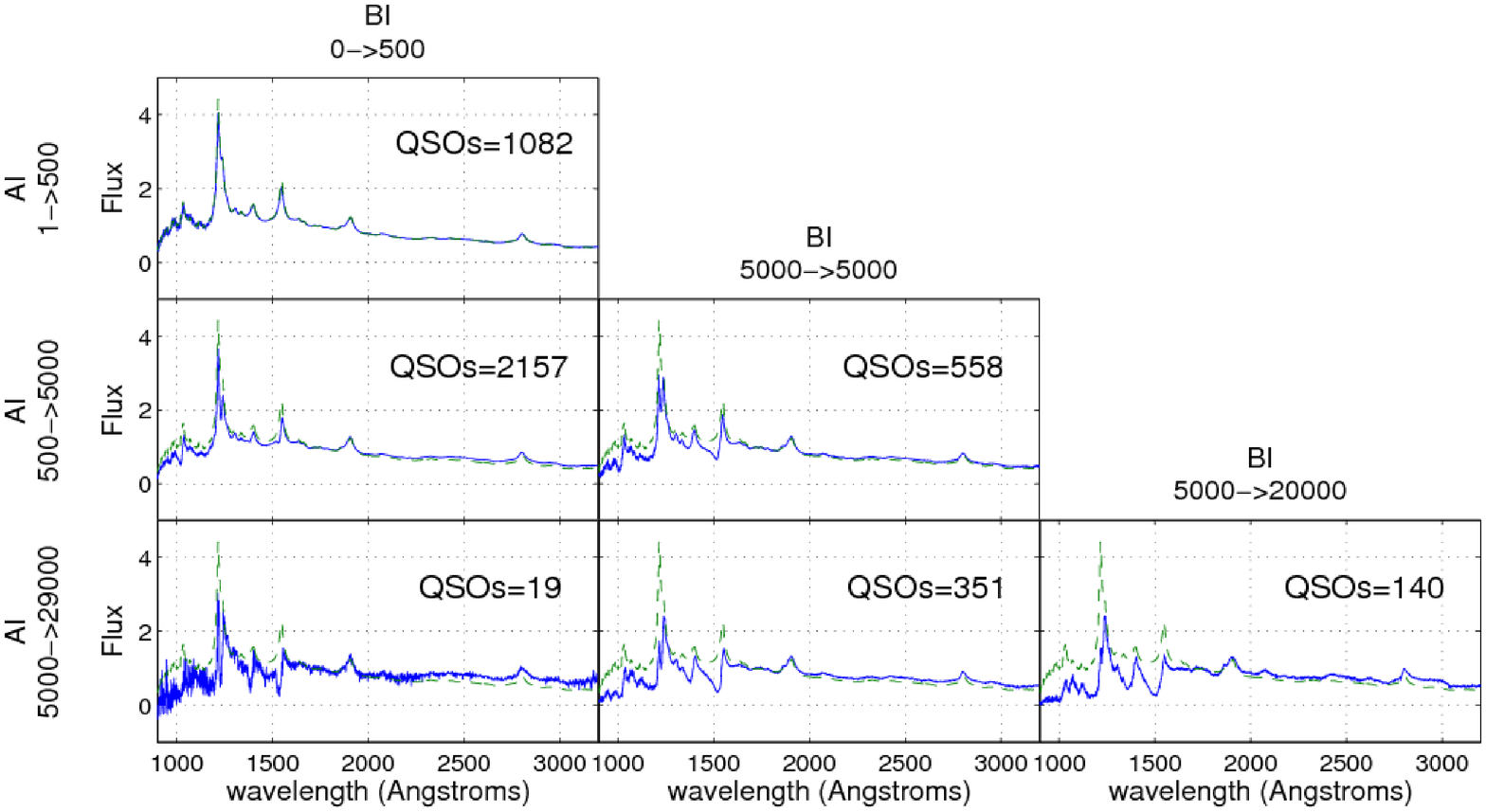}
\caption{Composites in various AI-BI ranges (blue line), and composites created from $AI=0$ and $BI=0$ objects. Reddening has not been taken into account.}
\label{fig:IX1}
\end{figure*}

Fig. \ref{fig:dr5_flow} shows a flow diagram of the steps
involved in creating the final DR5 BALQSO catalogue using the
hybrid-LVQ network.
Our LVQ-based method classifies 3,385 of the 28,421 QSOs ($11.91\% \mp 
0.21\%$) in our DR5 parent sample as BALQSOs and the catalogue can be 
found online\footnote{http://www.astro.soton.ac.uk/$\sim$simo}. It is reassuring to note that the LVQ classifications and the BI ones tend to agree for over $92\%$ of the objects. The ones which the methods disagree on are visually inspected for their classifications. Overall we find that $29\%$ of the $BI>0$ objects require visual inspection (848), whilst only $5\%$ (1224) of the $BI=0$ do. At first one might think that this is quite a high fraction for the $BI>0$ objects. Fig. \ref{fig:IX1} shows composites in various AI and BI bins. The top-left panel in the figure shows a composite made from 1082 objects with $0\geq BI \geq 500$ and $1 \geq AI \geq 500$. For comparison a composite produced with $BI=0$ and $AI=0$ has been overplotted with the dashed line. No signs of absorption are present, showing that QSOs with a $BI>0$ are not necessarily BALs (for more examples see \cite{knigge08}). The other panels in Fig. \ref{fig:IX1} show composites created for higher $AIs$ and $BIs$. The composite containing the most QSOs is the middle-left, which could also be considered as being the region in AI-BI space including the most borderline cases. This again helps explain the high fraction of objects which the BI and LVQ disagree.

\begin{figure*}
\centering \includegraphics[width=\textwidth, height=0.25\textheight]{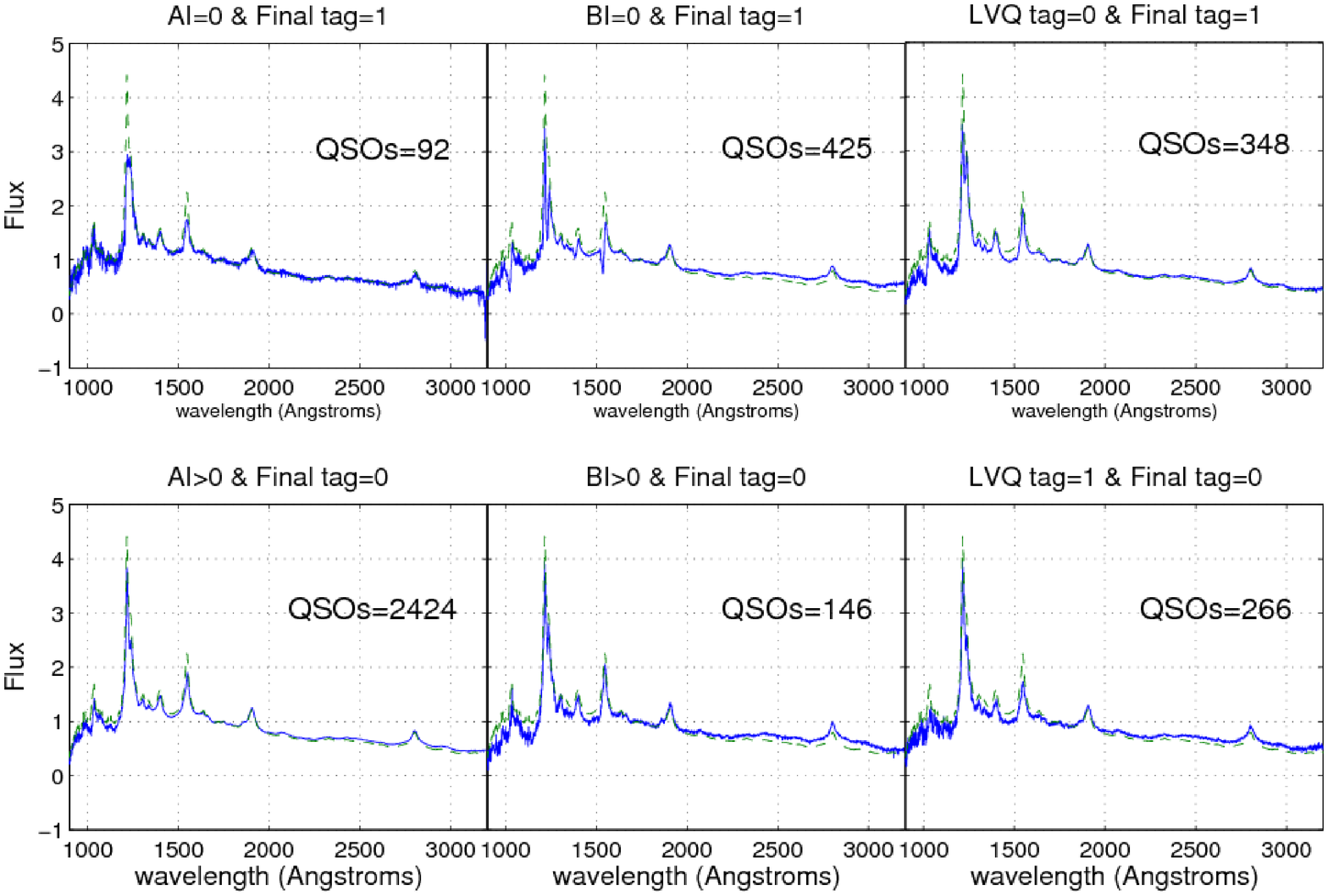}
\caption{Various composites created in order to check our hybrid-LVQ method (blue line), and composites created from $AI=0$ and $BI=0$ objects. LVQ tag=1 if LVQ classified the objects as BALQSOs, whilst Final tag==1 is set to objects which made it into our final BALQSO catalogue. Reddening has not been taken into account.}
\label{fig:IX2}
\end{figure*}

In order to explore our results in more detail we present the composites in Fig. \ref{fig:IX2}. The top row of panels show composites produced with QSOs which were finally tagged as BALQSOs, whilst the bottom row with non-BALQSO objects. As expected the top-centre composite displays a very narrow absorption, which will be missed by the BI calculation due to it's conservative definition. Next we consider the bottom-centre panel displaying the composite of objects with $BI>0$ but a non-BALQSO final tag. The spectrum is redder compared to the non-BALQSO composite, but has no clear signs of line absorption.
  
\section{Conclusions}

Compiling BAL quasar catalogues and determining the observed BALQSO
fraction is a challenging task. Most of the problem resides in the
ambiguity one encounters when attempting to classify individual
absorption features. After all, it is in the human nature to be
subjective, even if one tries the opposite. The judgment will be based
on previous examples of what is and what is not a BAL. This is why we
have used a hybrid method consisting of a simplified neural network
together with visual inspection to create a hopefully near-optimal BALQSO catalogue.

In \cite{knigge08} and this work we have shed light on many
classification problems when dealing with BALQSOs taken from
contemporary surveys. We showed that when the recently introduced
``absorption index'' (AI) is used to classify BALQSOs, the resulting
log AI distribution is clearly bimodal. Both modes containing
comparable amount of objects, but only the high-AI mode clearly being
associated with genuine BALQSOs. Moreover, in our previous paper, we
showed how even the traditional ``balnicity index'' (BI) produces
incomplete BALQSO samples. It is likely that due to the diverse nature
of observed BAL throughs, conventional metrics are no longer
appropriate given the large data volume increase in observed QSO
samples. Also it seems even more unfeasible to define new metrics in
order to deal with problems caused by the old ones. Here we have shown
how a hybrid algorithm can overcome these problems, taking into
account the increasing data volume gathered by contemporary
astronomical surveys.

The observed fraction is, however, still subject to serious selection
effects. In \cite{knigge08} we have explored these and corrected for
colour-, magnitude- and redshift-dependent selection biases on the DR3
dataset. After applying the corrections we reached the conclusion that
there is no compelling evidence for redshift evolution in the
intrinsic BALQSO fraction. 


\begin{theacknowledgments}
This work is supported by the Science and Technology Facilities Council. We thank the Sloan Digital Sky Survey for making their data publicly available.
\end{theacknowledgments}



\bibliographystyle{aipproc}   


\begin{thebibliography}{9}

\bibitem{foltz90}
C.~B. Foltz et al., ``The Third IBIS/ISGRI Soft Gamma-Ray Survey Catalog'', \emph{Bulletin of the American Astronomical Society}, \textbf{22}, pp. 806 (1990)

\bibitem{weymann91}
R.~J. Weymann et al., ``Continuum and Emission-Line Properties of Broad Absorption Line Quasars'', \emph{The Astrophysical Journal}, \textbf{373}, pp. 23--53 (1991)

\bibitem{reichard03b}
T.~A. Reichard et al., ``Comparisons of the emission-line and continuum properties of broad absorption line and normal quasi-stellar objects'', \emph{The Astronomical Journal}, \textbf{126}, pp. 2594--2607 (2003)

\bibitem{hewett03}
P.~C. Hewett and C.~B. Foltz., ``The Frequency and Radio Properties of Broad Absorption Line Quasars'', \emph{The Astronomical Journal}, \textbf{125}, pp. 1784--1794 (2003)

\bibitem{korista92}
K.~T. Korista et al., ``Hubble Space Telescope Faint Object Spectrograph and ground-based observations of the broad absorption line quasar 0226-1024'', \emph{The Astrophysical Journal}, \textbf{401}, pp. 529--542 (1992)

\bibitem{stocke92}
J.~T. Stocke et al., ``The radio properties of the broad-absorption-line QSOs'', \emph{The Astrophysical Journal}, \textbf{396}, pp. 487-503 (1992)

\bibitem{Tolea}
A.~ Tolea et al., ``Broad Absorption Line Quasars in the Early Data Release from the Sloan Digital Sky Survey'', \emph{The Astrophysical Journal}, \textbf{578}, pp. 31--35 (2002)

\bibitem{reichard03a}
T.~A. Reichard et al., ``A Catalog of Broad Absorption Line Quasars from the Sloan Digital Sky Survey Early Data Release'', \emph{The Astronomical Journal}, \textbf{125}, pp. 1711--1728 (2003)

\bibitem{knigge08}
C.~ Knigge et al., ``The Intrinsic Fraction of Broad Absorption Line Quasars'', \emph{Monthly Notices of the Royal Astronomical Society}, \textbf{386}, pp. 1426--1435 (2008)

\bibitem{trump06}
J.~R. Trump et al., ``A Catalog of Broad Absorption Line Quasars from the Sloan Digital Sky Survey Third Data Release'', \emph{The Astrophysical Journal Supplement Series}, \textbf{165}, pp. 1--18 (2006)

\bibitem{BIG_dr5}
J.~K. Adelman-McCarthy et al., ``The Fifth Data Release of the Sloan Digital Sky Survey'', \emph{The Astrophysical Journal Supplement Series}, \textbf{172}, pp. 634--644 (2007)

\bibitem{dr5}
D.~P. Schneider et al., ``The Sloan Digital Sky Survey Quasar Catalog. IV. Fifth Data Release'', \emph{The Astronomical Journal}, \textbf{134}, pp. 102--117 (2007)

\bibitem{north06}
M.~ North et al., ``A new sample of broad absorption-line quasars exhibiting the ghost of Lyman $\alpha$'', \emph{Monthly Notices of the Royal Astronomical Society}, \textbf{365}, pp. 1057--1066 (2006)

\bibitem{vanden01}
D.~E. Vanden Berk et al., ``Composite Quasar Spectra from the Sloan Digital Sky Survey'', \emph{The Astronomical Journal}, \textbf{122}, pp. 549--564 (2001)

\bibitem{kohonen01}
T.~ Kohonen, ``Self-organizing maps'', Springer series in information sciences, Berlin, Germany, 2001
 

\end{thebibliography}

\IfFileExists{\jobname.bbl}{}
 {\typeout{}
  \typeout{******************************************}
  \typeout{** Please run "bibtex \jobname" to optain}
  \typeout{** the bibliography and then re-run LaTeX}
  \typeout{** twice to fix the references!}
  \typeout{******************************************}
  \typeout{}
 }


\end{document}